\documentclass[usenatbib,twocolumn,useAMS]{mn2e}
\input{epsf}
\newcommand{\be}{\begin{equation}}
\newcommand{\ee}{\end{equation}}
\newcommand{\ba}{\begin{eqnarray}}
\newcommand{\ea}{\end{eqnarray}}
\newcommand{\bi}{\begin{itemize}}
\newcommand{\ei}{\end{itemize}}
\newcommand{\bfi}{\begin{figure}
\epsfxsize=9cm
\epsffile}
\newcommand{\efi}{\end{figure}}
\newcommand{\Msun}{M \odot}
\newcommand{\boldtheta}{\mbox{\boldmath$\theta$}}
\newcommand{\boldeps}{\mbox{\boldmath$\epsilon$}}

\newcommand{\mnras}{MNRAS}
\newcommand{\araa}{ARAA}
\newcommand{\apj}{ApJ}
\newcommand{\apjl}{\apj}

\newcommand{\prd}{PRD}
\newcommand{\prl}{PRL}
\newcommand{\nat}{Nature}
\newcommand{\physrep}{Physics Report}
\newcommand{\aj}{AJ}

\newcommand{\aap}{Astronomy and  Astrophysics}
\title[Lensing of 21cm Absorption ``Halos'' of First Galaxies]{Lensing
  of 21cm Absorption ``Halos'' of $z\sim$20-30 First Galaxies}  
\author[Pengjie Zhang, Zheng Zheng, \& Renyue Cen]
{Pengjie Zhang$^{1,2}$\footnotemark, Zheng Zheng$^{3,4}$\footnotemark
  \ and Renyue Cen$^{5}$\footnotemark
\\$^1$ Shanghai Astronomical Observatory, Chinese Academy of Sciences, 
 80 Nandan Road, Shanghai, China, 200030; pjzhang@shao.ac.cn
\\$^2$ Joint Institute for Galaxies and Cosmology (JOINGC) of SHAO and USTC
\\$^3$ Institute for Advanced Study, Einstein Drive, Princeton, NJ 08540, USA; 
 zhengz@ias.edu
\\$^4$ Hubble Fellow
\\$^5$ Princeton University Observatory, Princeton University,
Princeton, NJ 08544, USA; cen@astro.princeton.edu 
}

\begin{document}
\maketitle
\begin{abstract}
 Extended 21cm absorption  regions (dubbed ``21cm absorption halos'')
around first galaxies at $z\sim 30$ are likely the first
distinctive structures accessible to radio observations.  Though the
radio array capable of detecting and resolving them   must  have $\sim
200$ km$^2$ total collecting area, given
the great impact of such detections to the understanding of the
reionization process and cosmology, such radio survey would be extremely
profitable. As an example, we point out a potentially useful byproduct
of such survey. The  resolved 21cm absorption ``halos'',
likely close to spherical,  can serve as (almost) ideal sources for measuring
the {\it cosmic shear} and mapping the matter distribution to $z\sim 30$. 
We investigate the expected lensing signal and 
consider a variety of noise contributions on the shear measurement.
We find that S/N $\sim 1$ can be achieved for individual
``halos''. Given millions of 21cm absorption ``halos'' across the sky,
the total S/N will be   comparable to traditional shear measurement of
$\sim$$10^9$ galaxies at  $z\sim 1$. 
\end{abstract}


\section{Introduction}
Gravitational lensing (see, e.g., reviews of
\citealt{Bartelmann01,Refregier03})  directly probes the matter
distribution of the universe and is becoming one of the most important
probes of cosmology. 
Gravitational lensing distorts galaxy shape ({\it cosmic
  shear}),  changes galaxy number density ({\it cosmic magnification})
and induces mode-coupling in cosmic backgrounds. These observable
lensing effects enable several powerful methods to extract cosmological
information. Cosmic microwave background (CMB) lensing
\citep{Seljak99,Zaldarriaga99,Hu01,Hu02}, 21cm background lensing
\citep{Cooray04,Pen04,Zahn06,Mandel06} and cosmic magnification 
measurements of 21cm emitting galaxies \citep{Zhang05,Zhang06} are expected to
achieve high signal-to-noise ratio (S/N) in the near future.  On the other 
hand, cosmic shear measurements have achieved high S/N
(e.g., \citealt{Jarvis06,VanWaerbeke05,Hoekstra06}) and will be improved 
significantly by several ongoing or upcoming large lensing surveys.  

Traditional cosmic shear measurements are fundamentally limited by intrinsic 
ellipticities of galaxies. Intrinsic ellipticities have a dispersion $\sim
30\%$, much larger than the typical $1\%$ lensing signal for source galaxies 
at $z=1$.  Even in the best case that ellipticities of galaxies do not 
correlate, one still needs to average over several hundred galaxies to achieve 
S/N of one. Furthermore, source galaxies at $z s\ga 3$ are difficult to 
detect optically,  so  it is hard to map the matter distribution at redshifts 
beyond $2$ in optical lensing surveys. These two intrinsic problems can be 
overcome by the 21cm absorption regions around first galaxies 
\citep{Cen06}. These regions are of arc-minute size, or $\sim$
  Mpc in
comoving scale. Since their sizes are much larger than the nonlinear
scale at corresponding redshifts, they are believed 
to be very close to spherical \citep{Cen06}. In addition, since they 
lie at redshifts $z\sim 25$, the lensing signal is strong. 
These two intrinsic 
advantages make these regions nearly ideal targets for shear  measurement. 
Since these regions have distinctive structures, they are dubbed ``21cm
absorption halos'' or ``21cm halos''.  
We caution the readers that these 
21cm ``halos'' are not virialized dark matter halos. In fact, the density 
fluctuations in these ``halos'' are well in the linear regime. In this paper,
we study the requirement and application of the shear measurement of these 
21cm ``halos''.

Observations  of 21cm absorption ``halos'' are challenging and
 are way
 beyond the capability of the planned square
 kilometer array (SKA\footnote{http://www.skatelescope.org/}). 
However, mission capable of resolving these ``halos'' will be
 extremely profitable. The 21cm brightness temperature 
 can be simultaneously measured to high accuracy over a large fraction
 of  the sky  and a wide redshift 
 range, without extra cost. This will allow the measurement of related statistics, such as the
 brightness temperature power spectrum, bispectrum, etc., to
 unprecedented accuracy and significantly improve our understanding on
 reionization process and cosmology at high redshifts. The
 detection of  21cm absorption ``halos'' will have potentially great
 impact on cosmology too 
 \citep{Cen06}. Furthermore, the same mission can detect even smaller 21cm
 absorption regions around first stars, the so called Lyman-$\alpha$
 spheres (because 
it is mainly the Lyman-$\alpha$ scattering that couples the 21cm spin 
temperature to the kinetic temperature, leading to the 21cm absorption 
against the CMB, \citealt{Chen06}), through strong lensing of galaxy clusters
 \citep{Li07}. Given these potentially powerful applications, such
 mission deserve detailed investigation and observation efforts.  In
 this paper, we discuss a surprisingly valuable
 byproduct of such mission: mapping the matter distribution to $z\sim
 25$ through 
 weak lensing of 21cm absorption ``halos''.  

As we show in this paper, the ellipticity induced by various 
contaminations is very likely to be controlled to below $10\%$ for 21cm 
absorption regions considered here and we expect that shear measured from 
each such region would  achieve S/N$\sim 1$.  At $z\sim 25$ with $\Delta z=2$,
for a lower mass cut $\sim 7\times 10^7\Msun$, the total number of these  
21cm absorption ``halos'' is $\sim 10^6$ (Fig.~\ref{fig:n}).  The total S/N  
will be equal to that  
with $\sim 10^9$ $z\sim 1$ source galaxies and thus comparable to what would 
be provided by planned ambitious cosmic shear surveys such as
the Large Synoptic Survey Telescope\footnote{LSST:
  http://www.lsst.org/}.

Moreover, the lensing applications of 21cm absorption ``halos'' will
fill the gap of  
source redshifts between traditional cosmic shear measurements (source 
redshifts $z s\leq 3$), cosmic magnification of 21cm emitting galaxies 
($z s\sim 1$-$6$) and CMB lensing ($z s\simeq 1100$). They provide 
independent checks for 21cm background lensing ($z s\sim 10$---$30$).

Furthermore, lensing measurement to such high redshifts will provide a 
direct way of CMB delensing and improve the measurement of primordial 
inflationary gravitational waves \citep{Sigurdson05}. Lensing
reconstruction from  21cm absorption ``halos'' will significantly improve  our 
understanding of matter distribution to $z\sim 25$ and the nature of 
dark matter, dark energy, gravity and inflation.

The paper is structured as follows. After a brief introduction of the
21cm absorption ``halos'' of first galaxies in \S~2, we discuss
their detection prospect in \S~3. Then, in \S~4 we define the mean cosmic 
shear to be measured for these ``halos''. It would be ideal for cosmic 
shear measurement if the 21cm ``halos'' are intrinsically spherical. 
Therefore, in \S~5, by investigating several sources that may induce 
ellipticity of these ``halos'', we estimate the expected ellipticity and 
show that these regions are quite close to spherical. Given the expected 
noise, in \S~6 we show the range of lensing power spectrum that can be 
precisely measured.  Finally, we summarize and discuss the main results. 
Throughout the paper, we assume a spatially flat $\Lambda$CDM cosmology 
with matter density parameter $\Omega m=0.26$, baryon density parameter 
$\Omega b=0.044$, primordial power index $n s=0.95$, Hubble constant 
$H 0=100h=72 {\rm km~s^{-1}~Mpc^{-1}}$, and a fluctuation amplitude 
$\sigma 8=0.77$ on $8h^{-1}$Mpc scales, which is consistent with the 
Wilkinson Microwave Anisotropy Probe (WMAP) three-year results 
\citep{Spergel06}. For some plots, we also show cases with $\sigma 8=0.9$.

\section{21cm absorption ``halos'' of first galaxies}

Since the standard cold dark matter (CDM) cosmological model 
is by now accurately 
determined \citep{Spergel06},
one could make testable predictions for the first generation of galaxies,
expected to form within halos of mass $M=10^{5}$---$10^9 \Msun$ at
$z=20$---$40$ (Fig.~\ref{fig:n}).
The first galaxies are expected to be small and 
faint in terms of stellar optical light.
The very first galaxies
may have masses of $\sim 10^{5}\Msun$ (Peebles 1980),
limited by Jeans mass and molecular hydrogen cooling. 
However, star formation in minihalos with molecular cooling
might be quenched, when
hydrogen molecules are destroyed by Lyman-Werner photons.
In any case, because of the expected low star formation efficiency
in the minihalos, their 21 cm signals will be too weak to be
relevant even if they form stars.
However, as shown in \citet{Cen06}, with the coupling of the
21cm spin temperature to Lyman-$\alpha$ scattering,
some of the large first galaxies with total mass
$M\ge 10^{7.5}\Msun$ 
may each display an extended hydrogen 21cm absorption ``halo'' 
against the cosmic microwave background 
with a brightness temperature decrement of $\delta T=-(100$---$150)$~mK 
at a radius of $0.3$---$3.0$ Mpc (comoving),
corresponding to an angular size of $10$---$100$ arcseconds.
These large galaxies can cool by atomic processes.

\bfi{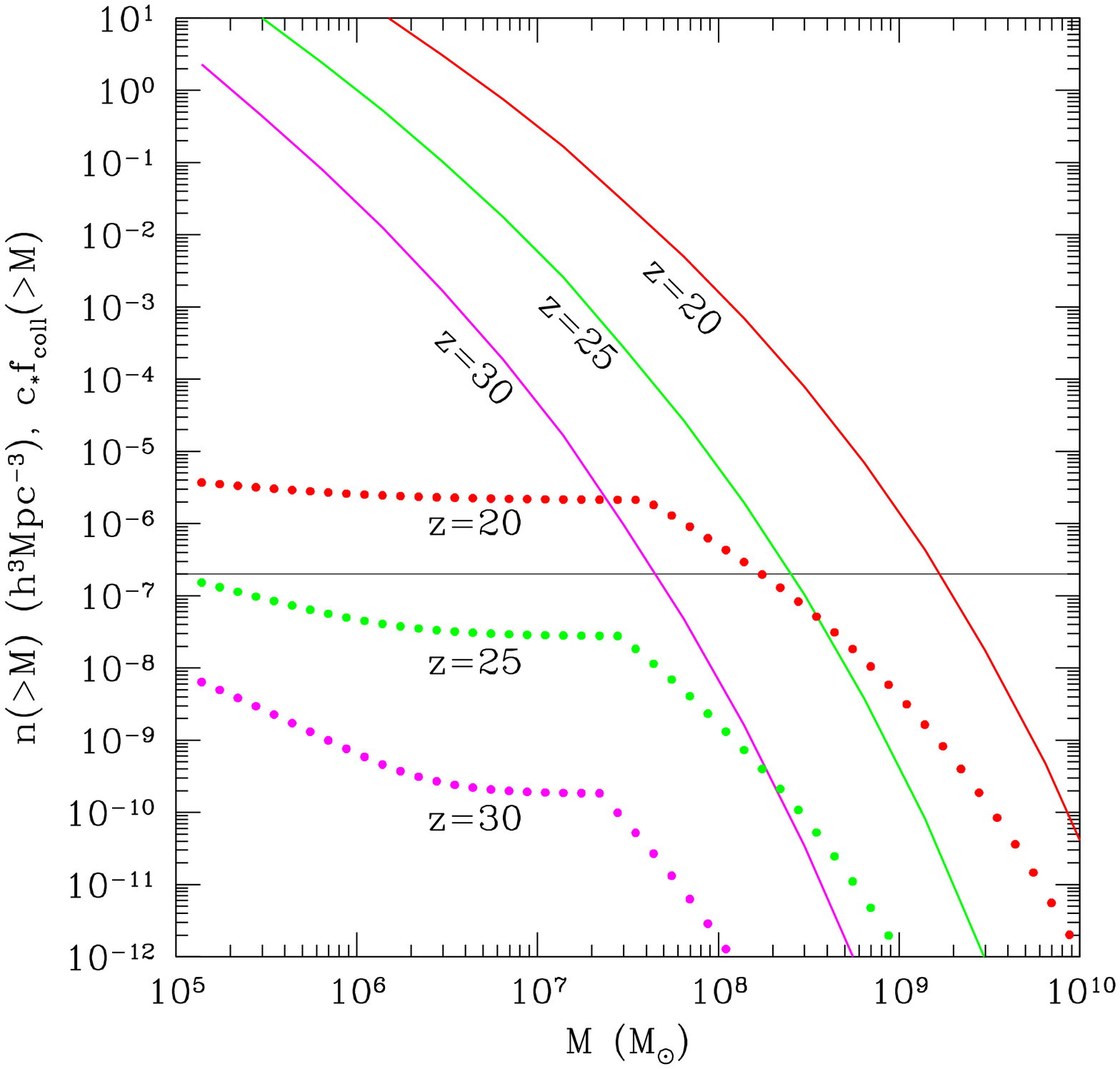}
\caption{Solid curves are cumulative halo mass functions at $z=$30, 25, and 20 
(bottom up), predicted by the Sheth-Tormen formula \citep{Sheth02}. Dotted
curves are products of the star formation efficiency $c *$ and 
the fraction $f {\rm coll}$ of matter that has collapsed to halos where
stars has formed at the three redshifts. The horizontal line represent a 
threshold value of $c *f {\rm coll}$, below which the heating of the IGM by
the X-ray background is not significant, with a temperature increment less
than $\sim$2K \citep{Cen06}. This figure is a reproduction of Fig.3 in 
\citet{Cen06} using the more accurate Sheth-Tormen mass function, assuming
$c *=0.2$ (0.001) for large halos (minihalos), and adopting cosmological 
parameters consistent with WMAP three-year results \citep{Spergel06},
$\Omega m=0.26$, $\Omega \Lambda=0.74$, $\Omega b=0.044$, $h=0.72$, 
$\sigma 8=0.77$ and $n s=0.95$. We focus on halos with mass greater
than $\sim 7 \times 10^7 M {\sun}$, which could harbor first galaxies capable 
of producing extended 21cm absorption regions against CMB \citep{Cen06}. 
\label{fig:n}}
\efi

The detection of these radio ``halos" around first galaxies will be 
extremely profitable, although admittedly difficult.  If and when a 21cm 
radio survey of first galaxies is carried out, some fundamental applications 
for cosmology and galaxy formation may be launched, 
as discussed  in Cen (2006).
Here, we recapitulate them.
First, it may yield 
direct information on star formation physics in first galaxies.  Second, it 
could provide a unique and sensitive probe of small-scale power in the 
standard cosmological model hence physics of dark matter and inflation,
by being able to, for example, constrain the primordial power index $n s$ to 
an accuracy of $\Delta n s=0.01$ at a high confidence level.  Constraints on 
the nature of dark matter particles, i.e., mass or temperature, or running of 
index could be still tighter.  Third, clustering of galaxies that may be 
computed with such a survey will provide an independent set of 
characterizations of potentially interesting features on large scales in the 
power spectrum including the baryonic oscillations, which may be compared to 
local measurements \citep{Eisenstein05} to shed light on gravitational growth 
and other involved processes from $z=30$ to $z=0$.  Finally, the 21cm 
absorption halos are expected to be highly spherical and trace the Hubble 
flow faithfully, and thus are ideal systems for an application of the 
Alcock-Paczy\'nski test \citep{Alcock79}.  Exceedingly accurate determinations 
of key cosmological parameters, in particular, the equation of state of the 
dark energy, may be finally realized.  As an example, it does not seem 
excessively difficult to determine the dark energy equation of state $w$ to 
an accuracy of $\Delta w\sim 0.01$, if $\Omega m$ has been determined to a 
high accuracy by different means.  If achieved, it may have profound 
ramifications pertaining dark energy and fundamental particle physics (e.g., 
\citealt{Upadhye05}).  This last property may also make them ideal sources 
for weak gravitational lensing measurements, as investigated here.

\section{Detectability}
The resolution and r.m.s. noise of a radio survey depend on the detailed 
array configuration. Without loss of generality, we focus on a configuration of 
$N\times N$ arrays homogeneously distributed over a square with side length 
$L$. The resolution is $\theta p=\sqrt{\pi \alpha^2/4} \lambda/L$, where
$\alpha\simeq 1.2$ for this specific configuration and $\lambda=0.21(1+z)$m 
is the redshifted wavelength of the 21cm line. Foreground contaminations 
at the low frequencies of concern here are overwhelming. 
The dominant one is  galactic
synchrotron emission, which scales as $\nu^\beta$ with
$\beta\sim$$-$2.55.  For concreteness, we adopt the
  brightness temperature at 
  $\nu$$\sim$54MHz (redshifted frequency of the 21cm  
lines at $z=25$) as  $T^{\rm syn} b\simeq 3000$ K. This number is
consistent with \citet{Keshet04} and \citet{Chen06}. However, brightness
temperature estimation at these low 
frequencies is quite uncertain (e.g., $T^{\rm syn} b$ adopted
by \citealt{Bowman06} is a factor of 2 higher, if scaled to $z=25$.).
 A factor of $2$ increase in $T^{\rm
  syn} b$ would require a factor of $4$ increase in integration time
or a factor of $2$ increase in the total collecting area, in order to
achieve the same S/N. 

Since the foreground is highly smooth
in frequency space and the 21cm signal has line features,  showing as
sharp fluctuations in the spectrum,  the mean foreground 
contamination can be efficiently removed pixel by pixel \citep{Wang06}. The 
residual noise  per resolution pixel (with area $\theta p^2$) caused by
photon number fluctuations in foreground and instrumental noise can be
worked out to be \citep{Thompson01}
\ba
\label{eqn:sigmaT}
\sigma T&\simeq&
\frac{T {\rm sys}}{\sqrt{\Delta \nu t}}\frac{4\sqrt{2}}{\pi\alpha^2f {\rm cover}}\\
&\simeq&
10 \ {\rm mK}\frac{T {\rm sys}}{3000 {\rm K}}
    \left(\frac{\Delta \nu}{300{\rm kHz}}\frac{t}{1{\rm year}}\right)^{-1/2}
    \left(\frac{f {\rm cover}}{14\%}\right)^{-1}\ .\nonumber 
\ea
Here, $f {\rm cover}=A {\rm total}/L^2=N^2A {\rm dish}/L^2$ and $A {\rm dish}$ 
is the collecting area of each dish. The bandwidth $\Delta \nu$ is related 
to the comoving separation $r$ by $\Delta \nu\simeq 46 {\rm kHz}
 [26/(1+z)]^{1/2} [r/h^{-1}{\rm Mpc}]$. The typical bandwidth 
$\Delta\nu=300{\rm kHz}$ corresponds to $\sim 6  h^{-1}{\rm Mpc}$ 
(comoving) at $z\sim 25$, which is about the diameter of the 21cm absorption 
``halos'' (see Fig.~\ref{fig:SN}). The integration time $t$ per pixel
is also the integration time per field of view (FOV). The system temperature 
$T {\rm sys}\simeq T b^{\rm syn}$ in our case.  We note that the noise 
estimate is much larger than that from the Poisson noise of photons. The 
reason is simple --- in the Rayleigh-Jeans part of the emission spectrum, the 
mean occupation number $\bar n$ of photons per quantum state is much greater 
than unity and the variance, which is $\bar n (\bar n +1)$, is super-Poisson.

\bfi{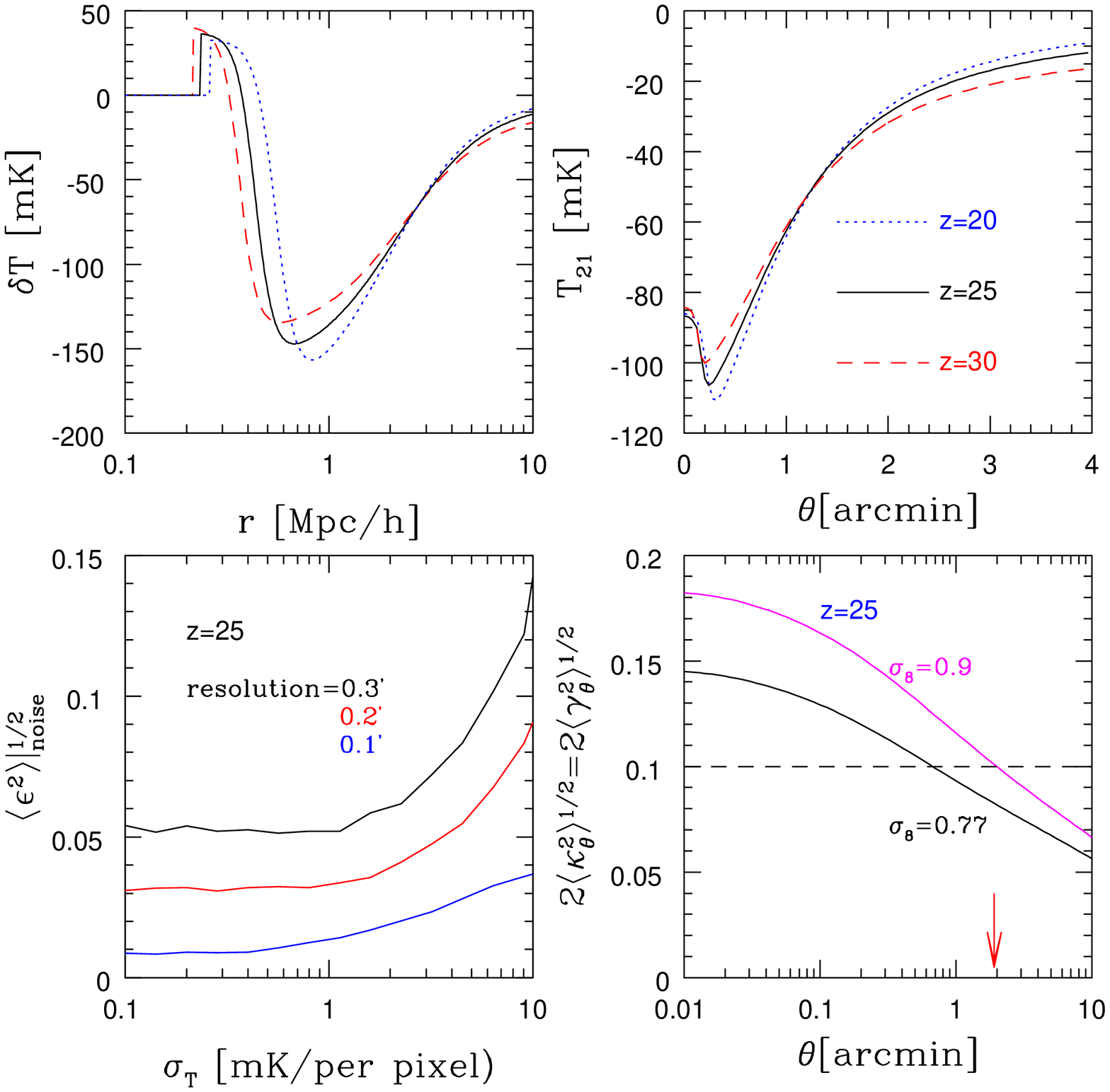}
\caption{
  Top left panel: the brightness temperature of a 21cm absorption ``halo'' 
  as a function of radius (in comoving unit), around a first galaxy in a 
  $10^8\Msun$ halo at $z=30,25,20$, respectively. The star formation 
  efficiency $c {*}=0.2$ is adopted.  Top right panel: the brightness 
  temperature averaged over $\pm 3 h^{-1}{\rm Mpc}$ (comoving) along the 
  line of sight. At these redshifts, 1 arcminute corresponds to $\simeq 
  2.4 h^{-1}{\rm Mpc}$ (comoving).  Bottom left panel: ellipticities caused 
  by random system noise (with an r.m.s. $\sigma T$ per resolution pixel). 
  See the text for details.  Bottom right panel: the lensing signal, $\kappa$ 
  or $\gamma$, smoothed over an area $\theta^2$. The arrow denotes the
  size of the 21cm absorption regions used for shear measurement. The
  dashed line is  
  roughly the expected noise level. \label{fig:SN}}
\efi

 With assumptions on the star formartion properties in the central
sources, the brightness temperature profile of the 21cm absorption ``halos''
can be computed (see \citealt{Cen06} for details). 
At $z=25$, 21cm absorbing  regions generated by first galaxies in 
$10^8 \Msun$ halos have angular size of several arc-minutes with signal 
$\la -50$mK (top panels, Fig.~\ref{fig:SN}). 
To reliably measure the shape of  
these ``halos'', a resolution of $\theta p\sim 0.3^\prime$ and a system 
temperature noise $\sigma T\sim 10$mK are required. For $\Delta \nu=300$kHz
and $f {\rm cover}=14\%$, this translates to $L\sim  40$ km, 
$A {\rm total}\sim 200$km$^2$, and an integration time of one year per FOV. 
Detecting such regions at $5$-$\sigma$ level without resolving them is 
relatively easy, which requires $L\sim 4$ km and $A {\rm total}\sim 2$ km$^2$. 
The FOV at these wavelengths is typically $\sim 10^3$ deg$^2$. 
To cover the whole sky, the total integration time of 
the order of  half a century is required.  Clearly, resolving
  these 21cm ``halos'' is not doable in the near future without major
  technological breakthrough or significant improvement in financial
  resources. Still,  theoretical merit of this study remains and could
be useful for mid and far future proposals on dark age science. 



The signal and size of the 21cm absorption ``halos'' do not have
strong dependence on the redshift (Fig. \ref{fig:SN}), until X-ray
heating becomes  important. However,  towards higher redshifts (e.g.,
$z=30$), thus lower frequency for observation, the foreground increases 
significantly.  Detecting and resolving these regions at $z=30$ would require 
more ambitious surveys. Furthermore, the number of halos drops by a factor 
of $\sim 100$ from $z=25$ to $z=30$ (Fig. \ref{fig:n}), which substantially 
reduces the significance of shear measurement at $z=30$. These factors make 
the measurement of shear for $z=30$ ``halos'' very unlikely. Towards  
lower redshifts, the shear measurement is also unlikely, but for a different 
reason. One necessary condition for the existence of the 21cm absorption 
feature around the first galaxies is that the intergalactic medium (IGM) is 
not significantly heated by the X-ray background. We recast the arguments 
in \citet{Cen06} by using the more accurate halo mass function \citep{Sheth02}
and adopting cosmological parameters consistent with WMAP three-year results 
\citep{Spergel06}. The results are shown in Figure~\ref{fig:n}, and we find 
that the above condition can be satisfied at $z\ga 23$. Therefore, we will 
focus our study for $z=25$.

\section{Cosmic shear}
To estimate the feasibility of shear measurement from 21cm absorption 
``halos'', we make several simplifications. We assume the dishes to 
distribute homogeneously in a square. This results in a homogeneous $u$-$v$ 
sampling and alleviates the work involved in map making. This allows us to 
follow the usual cosmic shear measurement strategy
and work in the real space. Real survey configurations often result in 
inhomogeneous $u$-$v$ sampling and  induced systematics in map  making
inevitably. However, \citet{Chang04} show that these systematics can be
corrected. Furthermore, the shapelet method, which  directly  
works in the $u$-$v$ space, is shown to be successful \citep{Chang04}
and should  be applied for real data. For these reasons, we
neglect these systematics  in our estimation.

Since the measurement noise is often large, we do not try tomography
in  frequency space for each 21cm absorption ``halos''. We average the measurements along the
line of sight over each  ``halo'' and  define the averaged
brightness temperature along the line of sight  
$T {21}=\int (\partial I/\partial T)\delta Td\nu/[\partial I/\partial
  T\Delta \nu]\simeq \int \delta T d\nu/\Delta \nu$.  Here,  $I$ is
the CMB intensity and $(\partial I/\partial T)\delta T$ is the
associated CMB intensity decrement caused by 21cm absorption. The
estimation shown here is by no means optimal and S/N 
could be improved by more advanced methods.

The two orthogonal components of the ellipticity \boldeps\ of a 21cm 
absorption ``halo'' are given by the estimator 
\be
\label{eqn:measure}
\epsilon 1=\frac{Q {11}-Q {22}}{Q {11}+Q {22}}; ~~~
\epsilon 2=\frac{2Q {12}}{Q {11}+Q {22}}
\ .
\ee
 The quadrapole moments $Q {ij}$ are defined by 
\be
Q {ij}\equiv\frac{\int q T[T {21}(\boldtheta)]\theta i\theta j d^2\theta}{\int
q T[T {21}(\boldtheta)]d^2\theta  } \ ,
\ee
where $q T(T {21})$ is a weighting function. We choose $q T(T {21})=T {21}$
when $T {21}<T {\rm th}$ and  $q T=0$ otherwise, and $T {\rm th}$ is a 
threshold of brightness temperature that defines the boundary of each region
used for shear measurement. The position \boldtheta\ is measured
with respect to the center with $i,j=1,2$ the two orthogonal directions. 
For a round object, \boldeps\ is induced by the cosmic shear and we have 
\be
\label{eqn:e}
\epsilon=\frac{2\gamma}{1-\kappa+\gamma^2/(1-\kappa)}\simeq
\frac{2\gamma}{1-\kappa}\equiv g\simeq 2\gamma\ .
\ee
Here, $\gamma$ and $\kappa$ are the lensing shear and convergence,
respectively.  With the presence of the
ellipticity $\epsilon N$ induced by  
other sources (see \S~5), we have $\epsilon\simeq
\epsilon N+2\gamma$. In equation~(\ref{eqn:e}), we have approximated the
reduced shear  $g\simeq 2\gamma$ since in general $\kappa\ll 1$. The
second order 
term $\kappa\gamma$ in $g$ has a $1$-$10\%$ contribution to the power
spectrum \citep{Dodelson06}, which slightly improves the
detectability. However, given the uncertainties in $\epsilon N$,
we neglect this higher order correction.

An implicit assumption of the estimator (eq.~\ref{eqn:e}) is that
cosmic shear does not vary across the integral area. This assumption
holds closely for galaxies, which are of arc-second size. The 21cm 
absorption ``halos'' are of arc-minute size and the cosmic shear does vary 
across these regions. In this case, the estimator (eq.~\ref{eqn:e}) measures 
the averaged cosmic shear. Higher order moments can be explored to extract 
the gradient of the cosmic shear (e.g., \citealt{Goldberg05}). The relatively 
large size of 21cm absorption ``halos'' could improve the S/N of such 
measurements. In this paper, we only discuss the mean shear signal of each 
21cm absorption ``halo''.

\section{Statistical errors}
Both the system temperature noise and inhomogeneities in neutral hydrogen 
density\footnote{Peculiar velocities are highly coherent across the 21cm 
absorption region, so it is unlikely to induce a non-negligible ellipticity.} 
and spin temperature induce effective ellipticity $\epsilon N$ to 21cm 
absorption regions. This $\epsilon N$ has expectation value zero and does 
not correlate over large distance. Analogous to galaxy intrinsic ellipticity, 
it only induces shot noise to the shear measurement. Since inhomogeneities 
and anisotropies in spin temperature, if exist, are mainly caused by the 
central source, they are uncorrelated with  inhomogeneities in neutral 
hydrogen density. The r.m.s. fluctuation in $\epsilon N$ then has three 
independent contributions:
\be
\label{eqn:totalerror}
\sigma^2 {\epsilon}\equiv\langle \epsilon N^2 \rangle=
  \langle \epsilon^2\rangle| {\rm noise}
+ \langle \epsilon^2\rangle| {\delta H}
+\langle \epsilon^2\rangle| {T s}\ ,
\ee
where $\langle \epsilon^2\rangle| {\rm noise}$, 
$\langle \epsilon^2\rangle| {\delta H}$, and 
$\langle \epsilon^2\rangle| {T s}$ are ellipticity fluctuations caused by
the system temperature noise, the inhomogeneity in the neutral hydrogen
distribution, and the anisotropic distribution of the spin temperature
around the first galaxies, respectively. 

\subsection{Ellipticity induced by the noise in system temperature}
The distribution of the system temperature noise is Gaussian with an r.m.s. 
$\sigma T$ given by equation~(\ref{eqn:sigmaT}). To estimate the fluctuation
$\langle \epsilon^2\rangle| {\rm noise}$ in the ellipticity induced by this
noise, we generate a series of realizations of observations by adding 
Gaussian noise with an r.m.s. fluctuation $\sigma T$ to the otherwise 
spherically distributed signal and measure the ellipticity using 
equation~(\ref{eqn:measure}). The result is shown in the bottom left panel 
of Figure~\ref{fig:SN}. For $\sigma T\sim 10$mK and resolution $0.3^\prime$, 
the induced ellipticity is $\sim 10\%$ for $z=25$. To have
$\langle \epsilon^2\rangle| {\rm noise}^{1/2}<0.1$, $\sigma T<7$ mK is
required. This translates to a total collecting area $\ga 300$ km$^2$
for $L=40$ km.  The induced ellipticities at other redshifts
(e.g., $z=20$ or 30) are  similar, due to their similar signals
(Fig. \ref{fig:SN}).

We notice that 
even for $\sigma T\rightarrow 0$, $\langle \epsilon^2\rangle| {\rm noise}$
does not vanish, rather it reaches a plateau instead. This is caused by the 
limited spatial resolution. With a finite resolution, one can not exactly 
identify the physical center of each regions, and any shift from the 
physical center would result in an effective ellipticity. 
As shown in the bottom left panel of Figure~\ref{fig:SN}, with fixed 
$\sigma T$, one can indeed reduce $\langle \epsilon^2\rangle| {\rm noise}$ 
significantly by improving the angular resolution $\theta p$. However, in 
reality, the requirement to improve resolution is often in conflict with 
the requirement to reduce the noise.  To improve the angular resolution, 
longer base line is required ($\theta p\propto L^{-1}$). If the total 
collecting area is fixed, $f {\rm cover}$ would decrease ($f {\rm cover}
\propto L^{-2}$). This would cause the noise $\sigma T$ (per resolution 
pixel) to increase ($\sigma T\propto f^{-1} {\rm cover}\propto L^2\propto
\theta p^{-2}$) and $\langle \epsilon^2\rangle| {\rm noise}$ to
increase as well (Fig. \ref{fig:SN}). Therefore, improving angular
resolution does not necessarily reduce errors in shear measurement,
unless the total collecting area is increased accordingly. Optimal
surveys may balance between the angular resolution and $\sigma T$ to
reduce $\langle \epsilon^2\rangle| {\rm noise}^{1/2}$ to a level well
below the intrinsic ellipticity induced by the spin temperature anisotropy
(\S \ref{subsec:spin}). 

\subsection{Ellipticity induced by density inhomogeneities}

The fluctuation $\langle \epsilon^2\rangle| {\delta H}$ in the ellipticity 
induced by inhomogeneities in neutral hydrogen density can be estimated as 
\ba
\label{eqn:density}
\langle \epsilon^2 \rangle {\delta H}& \simeq& \left[\int q T\theta^2
  d^2\theta\right]^{-2} \times \int q T(T {21})q T(T {21}^\prime) 
  w(\boldtheta-\boldtheta^\prime) \nonumber\\
&&\times
        [(\theta 1^2-\theta 2^2)({\theta 1^\prime}^2-{\theta 2^\prime}^2)+4\theta 1\theta 2\theta 1^\prime\theta^\prime 2]d^2\theta d^2\theta^\prime \ . 
\ea
Here, $w(\theta)$ is the angular correlation function of the neutral
hydrogen over-density. In this expression, we have neglected fluctuations in 
the denominator $Q {11}+Q {22}$ (eq.~\ref{eqn:measure}) since they only cause 
second order effect, 
due to non-vanishing $Q {11}+Q {22}$.  For a flat $w(\theta)$, $\langle
  \epsilon^2\rangle| {\delta H}$ vanishes due to the geometry terms in
equation~(\ref{eqn:density}). In reality, $w(\theta)$ varies slowly. So 
$w(\theta)$ does not contribute much to the integral in 
equation~(\ref{eqn:density}), although itself can reach 
$\sim 10^{-2}$ at sub arc-minute scale.
For a 21cm absorption ``halo'' of a galaxy in a $10^8 \Msun$ halo at 
$z=25$, $\langle \epsilon^2\rangle| {\delta H}< 10^{-4} \ll
4\langle \gamma^2\rangle$ for $T {\rm th}\sim -50$mK. 
Therefore, this source of error seems to be negligible. 

First galaxies are more likely to reside in high density regions and
one may think that this strong formation bias invalidates the above
conclusion, which is based on the ensemble average of the correlation
function. As the shear  measurement only depends on the variation 
in the local density field, to account for the environmental effect, one 
should replace $w(\theta)$ in equation~(\ref{eqn:density}) with the
conditional correlation function  $w(\theta|\langle \delta H
\rangle)$, where $\langle \delta H \rangle$ is  the overdensity
averaged over the 21cm absorption regions for shear measurement. Being
several Mpc/$h$ in size, these regions are much larger than the
nonlinear scale or filament size at $z\sim 25$,  which implies that 
$\langle \delta H \rangle\ll 1$. Hence we do not expect that 
$w(\theta|\langle \delta H \rangle)\gg w(\theta)$. On the other hand, 
even if $w(\theta|\langle \delta H \rangle)$ could be 10 times larger 
than $w(\theta)$, the induced $\langle \epsilon^2 \rangle {\delta H}^{1/2}$ 
would be still less than $3\%$. 
Based on these arguments, we conclude that the ellipticity induced by 
inhomogeneities in neutral hydrogen density, if not negligible, is at
most sub-dominant to other error sources.   

\subsection{Ellipticity induced by spin temperature inhomogeneities}
\label{subsec:spin}
The fluctuation $\langle \epsilon^2\rangle| {T s}$ in the ellipticity induced 
by anisotropies in the spin temperature distribution is most uncertain. 
The spin temperature is a
weighted average of the CMB temperature and the gas kinetic temperature.
For the temperature range in our study, the CMB temperature is perfectly 
homogeneous. The coupling coefficients of the spin temperature to the gas 
temperature are determined by the spin-changing collisional rate and the
Lyman-$\alpha$ scattering rate \citep{Wouthuysen52,Field59}. On scales
we consider, the collisional rate should be isotropic because of 
isotropic distributions of the gas temperature and density. However, the 
spatial distribution of the Lyman-$\alpha$ scattering rate highly
depends on the origin and propagation of Lyman-$\alpha$ photons, which could
introduce considerable anisotropies in the spin temperature distribution.

\citet{Cen06} considers Lyman-$\alpha$ photons originated from the galaxy at 
the center of each halo. Photons blueward Lyman-$\alpha$ from the 
continuum of Population III stars in the galaxy redshift to the 
Lyman-$\alpha$ line-center frequency because of the Hubble expansion. 
At the radius where a blue photon redshifts to Lyman-$\alpha$ frequency,
the photon is continuously scattered by neutral hydrogen atoms in the IGM
and has little spatial diffusion. The photon would not fly freely until it 
encounters an atom that has the right velocity to make a large jump in its 
frequency. The escape of photons slightly blueward the Lyman-$\alpha$ 
line-center frequency from the central galaxy is affected by the neutral 
hydrogen distribution near the center. If the neutral hydrogen column 
density in the innermost 1kpc of the halo is above $10^{17}{\rm cm}^2$ (which 
is highly possible) and the column density distribution is not isotropic 
(e.g., caused by the geometry of the star forming region or the galactic 
wind), it would lead to anisotropic distributions of the Lyman-$\alpha$ 
scattering rate and the spin temperature at large radii. 

However, before a photon reaches the radius where it redshifts to 
Lyman-$\alpha$ and becomes strongly scattered, there is a probability for it 
to be scattered and thus change its propagation direction. The place where 
the photon encounters a tremendous number of scatterings with little spatial 
diffusion is then no longer along the initial direction that the photon 
escapes from the central galaxy. Therefore, a small number of scatterings 
encountered by photons before they reach the region where strong scatterings 
happen tend to make the distribution of the overall scattering rate more 
isotropic than the initial angular distribution of the photons from the 
central galaxy.

To quantify such an effect of isotropization, we simulate the Lyman-$\alpha$ 
scattering process around the galaxy using a Monte Carlo code 
\citep{Zheng02}. Photons near Lyman-$\alpha$ frequency are launched from 
the center along a cone with a given open angle to represent the anisotropic 
initial distribution. Every scattering of a photon is followed until the 
photon escapes to infinity. The scattering rate at each position in the IGM 
around the galaxy is then obtained with its normalization set by the 
luminosity of the central galaxy around Lyman-$\alpha$ frequency (see 
\citealt{Cen06}). We calculate the resultant spin temperature
distribution and find that with a threshold $T {\rm th}=-50{\rm mK}$
of the 21cm absorption region, the ellipticity fluctuation 
$\langle \epsilon^2\rangle| {T s}^{1/2}$ introduced by the 
initial anisotropic Lyman-$\alpha$ emission 
are 0.21, 0.20, 0.16, 0.14, 0.08, 
and 0.03 for open angles of 5$^\circ$, 15$^\circ$, 30$^\circ$, 45$^\circ$, 
60$^\circ$, and 75$^\circ$, respectively. 
 We may expect that galaxy
formation at such high 
redshifts is quite irregular and thus the formation of a well
defined disk is unlikely. If this is the case, it may be unlikely that
the effective open angles are much smaller than $\sim$30$^\circ$.
But even in such extreme cases, a small number of initial scatterings are 
quite efficient in isotropizing the 
final scattering rate distribution by making most resonant scatterings
happen at a position along a direction deviating from the initial one.

The situation may be even more optimistic than the above estimates.
Besides photons originally emitted slightly blueward the Lyman-$\alpha$
frequency that redshift to the Lyman-$\alpha$ frequency, there are other
sources of Lyman-$\alpha$ photons that can contribute to the pumping rate
of the 21cm line.
Photons originally emitted between Lyman-$\gamma$ and Lyman limit can
redshift into one of the higher Lyman series resonance, and, after a few
scatterings, cascade to locally produce Lyman-$\alpha$ photons near the
line center (\citealt{Chuzhoy06a,Chuzhoy06b}). Owing to the lower
cross-section of the photons between Lyman-$\gamma$ and Lyman limit, they
are expected to escape from the central galaxy more isotropically than
those slightly blueward Lyman-$\alpha$. The small number of scatterings
before they cascade to produce Lyman-$\alpha$ photons would further
isotropize the distribution. Another source of Lyman-$\alpha$ photons is
from soft X-ray photons. \citet{Chen06} discuss 21cm absorption halos
around first stars and argue that soft X-ray photons from Population III
stars play a significant role in (collisionally) generating Lyman-$\alpha$
photons. These Lyman-$\alpha$ photons produced locally in the inner IGM
region surrounding the galaxy are likely to be largely isotropic because
of the largely isotropic escape of soft X-ray photons from the galaxy.

Based on the above investigation, we conclude that the 
ellipticity induced by the anisotropies in the spin temperature is likely
at the level of $\le 0.1$ but may be much smaller. 
Depending on the nature
of these objects and observation configurations, either the noise
induced ellipticity or the spin temperature induced ellipticity can
dominate the error budget of shear measurement. 
The overall induced
ellipticity is thus expected to be $\le 10\%$ but could be smaller.

\bfi{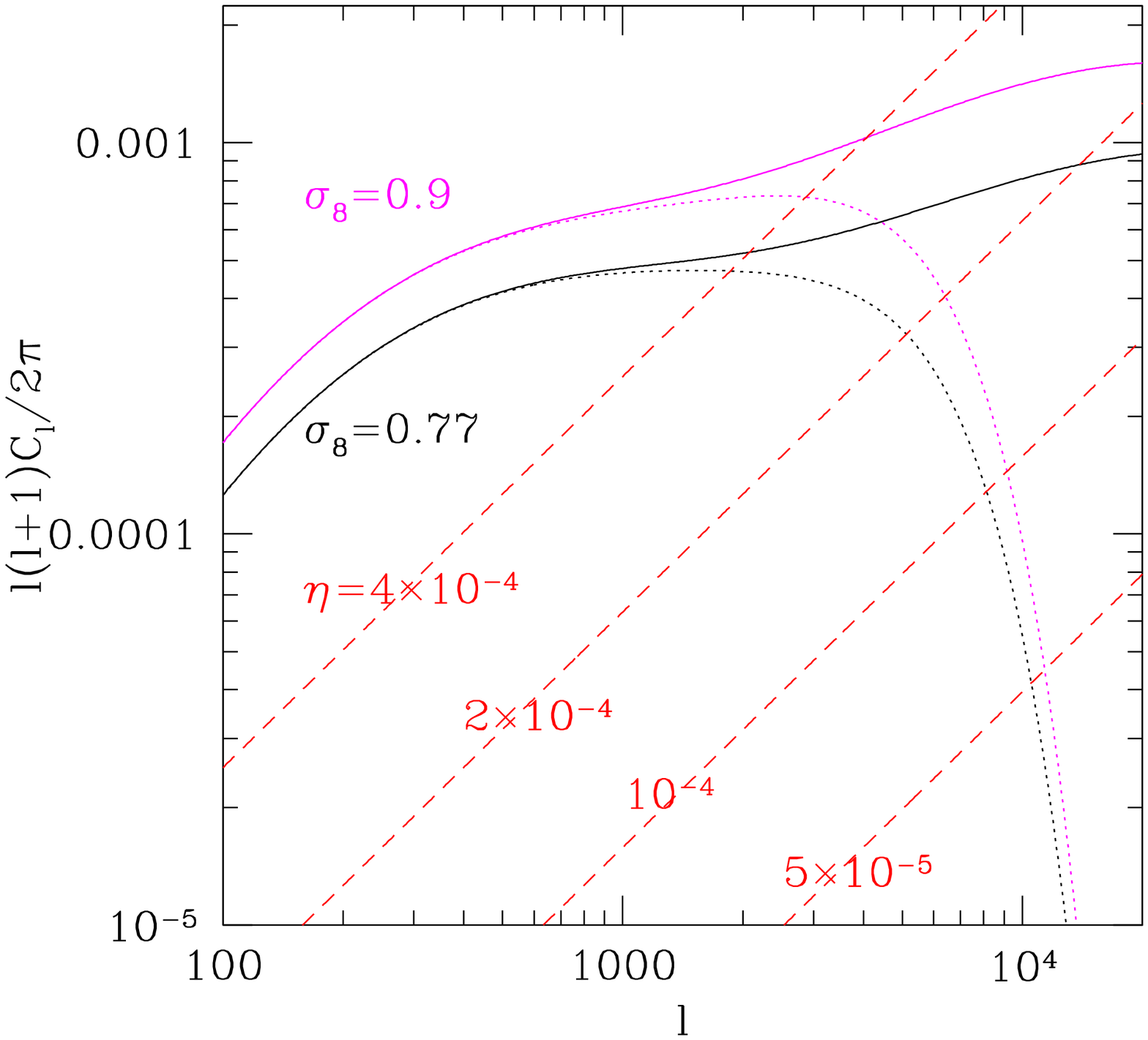}
\caption{Error forecast for lensing power spectrum from shear measurement of
$z\sim$25 21cm absorption ``halos''. The solid curves are the
  lensing power spectra (without smoothing) at $z=25$ for $\sigma 8=0.77$ and
$\sigma 8=0.9$, respectively. Dashed lines are statistical error
  caused by shot noises as a function of 
$\eta\equiv \sigma {\epsilon}/N g^{1/2}$, which is $\sim 10^{-4}$ for
typical values of $\sigma {\epsilon}\sim 0.1$ and $N g\sim 10^6$ with 
$\Delta l=0.1l$ adopted. The quadrupole moment shear
  estimator measures the mean shear averaged over an 
aperture and allows the reconstruction of lensing at $l\la 5000$ (dot
  lines, with  smoothing area 4 arcmin$^2$). Small scale lensing power
  $l\ga 5000$ 
  can not be measured  by the quadrupole moments shown here. However,
  it can be recovered by higher order moments such as the octopole
  moments. Eventually, one can recover all 
  lensing information up to the limit of shot noise.  \label{fig:error}}  
\efi
\section{Lensing measurements}
The lensing signal on arc-minute scales at $z=25$ is $2\kappa \sim
10\%$ (lower right panel, Fig.~\ref{fig:SN}). From the previous
section, the expected noise is likely $\la 
10\%$. We then expect that the S/N of the shear measurement for each 21cm
absorption ``halo'' is $\sim 1$. This S/N is impressive, comparing to
S/N$\sim 1/30$ of 
the conventional shear measurement of optical galaxies at $z=1$. 

Combining cosmic shear measurements
of all ``halos'', one can measure the
lensing power spectrum $C l$, where $l$ is the multipole. The 
statistical error in $C l$, assuming Gaussianity, is
\be
\Delta C l=\sqrt{\frac{2}{(2l+1)\Delta lf {\rm sky}}}\left(C l+\frac{4\pi f {\rm
      sky}\sigma {\epsilon}^2}{N g}\right) \ .
\ee
Here, the first term is the cosmic variance and the second term is the shot
noise, $f {\rm sky}$ is the sky coverage and $\Delta l$ is the size of the 
$l$ bin, $N g$ is the number of 21cm absorption ``halos''. Since we measure 
the shear averaged over the size of each 21cm absorption ``halo'', one should 
replace $C l$ with the smoothed $\tilde{C} l=C lW^2 {\theta}(l)$, with 
$W \theta$ 
being the Fourier transform of the window function describing each 21cm 
absorption region used for lensing measurement.  

For a lower mass cut $7\times 10^7 \Msun$, there are about $10^6$ first 
galaxies and thus $10^6$ 21cm absorption ``halos'' across the sky at 
$24<z<26$.  The shot noise power spectrum is proportional to the square of 
$\eta\equiv \sigma {\epsilon}/N g^{1/2}$. With $\sigma {\epsilon}\sim 0.1$ 
and $N g\sim 10^6$, the typical value of $\eta$ is $10^{-4}$, for which
precision measurement can be done up to  $l\sim 5000$ (several arc-minute 
scale; Fig.\ref{fig:error}). 

The lensing reconstruction from 21cm ``halos'' can be further improved.
Figure~\ref{fig:error} shows that there is significant  power at small 
scales ($l\ga 5000$).  Since the quadrupole moment shear estimator only 
measures the averaged shear, the smoothing effect caused by the arcminute 
size of the smoothing area does not allow the extraction of small scale 
lensing information. By measuring higher order moments, however, smaller 
scale lensing information can be recovered.  In the literature,  the octopole 
moments have been proposed to measure the gradients of shear, or equivalently, 
the local shear (e.g., \citealt{Goldberg02}). The relative 
large size of 21cm absorption ``halos'' will make this measurement
much easier than that of $z\sim 1$ galaxies. Eventually, by measuring
all necessary moments, one can recover all lensing information up to
the limit of shot noise. This will allow the measurement of the lensing
signal to $l\sim 10^4$ (Fig.~\ref{fig:error}).

In the above estimation, we have assumed that intrinsic
  ellipticities do not correlate over relevant scales and thus only
  contribute to shot noise. This is obviously true for that induced by local
  processes discussed in previous sections, such as local density
  inhomogeneities and spin temperature anisotropy caused by the
  central sources. Could a fluctuating large scale field coupled to the
  brightness temperature, such as the  fluctuating soft X-ray background
  (e.g. \citealt{Pritchard07}), invalidate the above assumption? The answer is
  probably no. The modes capable of inducing ellipticities must be
  smaller than the $\sim$ Mpc size of 21cm absorption ``halos''. On the
  other hand, the
  modes capable of inducing correlations in intrinsic ellipticities
  are of size $\sim 2\times 10^4/l \ h^{-1}$ Mpc. Thus a fluctuating
  background can only induce correlations in intrinsic
  ellipticities for those modes of $l>10^4$, which are of little
  relevance to this paper.

  In the directions along the line of sight and perpendicular to the line 
  of sight, 21cm absorption halos may appear different, due to the 
  evolution of these halos and the light cone effect, the 
  redshift distortion, the beam  and foreground (which are frequency
  and thus redshift 
  dependent), etc. However, these physics do not induce asymmetry in
  the  2D plane perpendicular to the line of sight. Since we only use
  the shape distortion in the same 2D plane to measure cosmic shear,
  these physics do not induce systematics in the shear measurement. 

Thus, with reasonable estimation of the 21cm absorption ``halo''
signal at high redshifts,  rather conservative estimates for the
intrinsic ellipticities caused by initial anisotropic Lyman-$\alpha$
emission from galaxies and reasonable estimates on the  level of
foreground induced ellipticity, the weak lensing application of these
``halos'' are very promising. The combined lensing S/N of $\sim 10^6$
21cm ``halos'' at $z=24-26$ is comparable to those with traditional
cosmic shear  
measurements, with cosmic magnification of 21cm emitting galaxies, and with 
CMB lensing and 21cm background lensing.

\section{Summary and Discussion}

In this paper, we investigate the potential of using 21cm absorption 
``halos'' of first galaxies at $z\sim 25$ as background sources for 
gravitational weak lensing reconstruction. We show that the accuracy 
obtained using this method can be comparable, and may be superior, to
some of the conventional methods.  It is potentially very rewarding 
scientifically to detect and survey the first galaxies using 
future 21cm radio experiments for this and other applications. 

There are several major uncertainties that may affect the results in this
paper. The first one is the star formation efficiency $c *$ in large halos
at $z\sim25$. We have adopted $c *=0.2$ in the calculation. A smaller 
$c *$ would reduce the signal and size of the 21cm ``halos'' and thus make
the observations and shear measurement more difficult. We have checked
that adopting $c *=0.1$ doubles the r.m.s. of the ellipticity induced 
by the system noise for the same noise level $\sigma T=7$ mK. Secondly,
there are uncertainties in the hard X-ray heating, which depends on the 
energy extraction efficiency from black holes and the fraction of the 
released energy in the form of hard X-rays (see \citealt{Cen06} for more 
details). If the real values are lower than what are adopted in our 
calculation, the X-ray heating would be less important and the 21cm 
absorption ``halos'' may form at redshifts lower than $z=25$. As a 
consequence, we would have much more observable 21cm absorption ``halos'' 
and the statistical errors of the lensing measurement would be significantly 
improved.  If the X-ray heating is more efficient than we assume, the 
detection of 21cm absorption ``halos'' would be more demanding and it 
becomes more difficult to use these ``halos'' for the lensing application. 
To reduce the above two types of uncertainties, we have to advance our 
understanding of the formation of the first objects. On the other hand, 
the observation (or even null observation) of the 21cm absorption ``halos'' 
would lead to valuable constraints on star formation and X-ray heating at 
$z\sim 20$---30.  Lastly, we have neglected any errors in the shear 
measurement induced by map making. Although in principle these errors can be 
corrected \citep{Chang04}, it is not clear how well the correction is for 
unprecedented radio arrays required for 21cm absorption ``halo'' observations. 
The discussion of such residual errors is certainly beyond the scope of this 
paper. 

While we focus here on the weak lensing application of the 21cm 
absorption ``halos'', because of their arc-minute size and high redshifts, 
these ``halos'' also have interesting strong lensing signatures, which is 
discussed in \citet{Li07}.

\section*{Acknowledgments}
We thank Gary Bernstein, Xuelei Chen, Leonid Chuzhoy, Bhuvnesh Jain, Yipeng 
Jing, Jordi Miralda-Escud\'e and Zhiqiang Shen for helpful discussions.  We 
thank Ue-Li Pen and Olivier Zahn for valuable information on observations. 
PJZ is supported  
by the One-Hundred-Talent Program of Chinese academy of science and the NSFC 
grants (No.  10543004, 10533030). ZZ acknowledges the support of NASA through 
Hubble Fellowship grant HF-01181.01-A awarded by the Space Telescope Science 
Institute, which is operated by the Association of Universities for Research
in Astronomy, Inc., for NASA, under contract NAS 5-26555. RC is supported in 
part by grants AST-0407176 and NNG06GI09G.

\end{document}